\documentclass[%
showkeys,
showpacs,
reprint,
superscriptaddress,
nofootinbib,
nobibnotes,
 amsmath,amssymb,
prb,
]{revtex4-2}
\usepackage{amssymb}
\usepackage{color}
\usepackage{graphicx}% Include figure files
\usepackage{dcolumn}% Align table columns on decimal point
\usepackage{bm}% bold math
\usepackage{hyperref}
\usepackage{breakurl}
\begin{document}

\title{Evaluation of gamma- and electrons- irradiation effects in organic, inorganic and biological Substances\\ \emph{A phenomenological study}}
\thanks{Presented on the 9th Environmental Physics Conference, 26-29 March, 2021, Aswan-Luxor, Egypt.}

\author{Elsayed~K.~Elmaghraby}\email[Corresponding Author: ]{e.m.k.elmaghraby@gmail.com}
\email{elsayed.elmaghraby@eaea.org.eg}
\address{Experimental Nuclear Physics Department, Nuclear Research Center, Egyptian Atomic Energy Authority, Cairo 13759, Egypt.}

\author{Mohamed Bahaaeldin-Afifi}
\address{Medical physics department, Minia Oncology Center, ministry of health and population, Al-Minia, Egypt.}

\date{\today}% It is always \today, today,
 % but any date may be explicitly specified

\begin{abstract}
The pattern of radiation energy deposition in substances at the microscopic level of lattice, molecule size, or the cell's nucleus is not uniform. The energy of radiation is transferred to the substance medium in the form of discrete, time-dependent, spatially correlated events with excitations/ionizations are the processes involved. The response of material on the macroscopic level to radiation effects depends on the microscopic pattern of energy deposition. A mathematical model that combines the specific number of sites available for the interaction of radiation and the detected signals of the property was proposed and discussed. This model emphasizes the phenomenon of log-dose response for moderate amount of nuclear radiation affects the material, especially detector materials. A parameter ($\alpha _{\nu }^{\bullet }$) was adopted to represent the remaining fraction of sites that were affected in the material by one unit dose at which the damage/change/modification of its properties occurs. The recovery factor of the effect and the delayed retardation enhancement factor are included in the model.
\end{abstract}

\maketitle

\section{Introduction}

More than twelve decades have been elapsed since the first discovery of radioactivity. During this time, numerous natural or artificial sources of radiation have been discovered or invented; beginning from radium and Roentgen x-ray machine to neutron generators and synchrotron radiation. The applications, as well as hazards of these types of radiation, depending on their physical nature; including charge, mass, and energy. Radiation scientists spent most of their time studying sources, action and interaction, applicability for public well-being, and ways to avoid its hazards [1-3]. At the atomic scale - which is best studied in the gaseous phase - the interaction of radiation with atoms had led to the emergence of quantum mechanics [4]. In this situation, the interaction regimes are the energy transfer by coulomb's excitation, photo-electron emission, and Compton's scattering, or matter production and annihilation (We recall these as Primary Interactions PI). On the other extreme, when all the energy and mass of the radiation are absorbed by the material, the bulk material may change its macroscopic properties [5], at least its temperature, as a consequence of being subjected to energy transfer. One of the main consequences of the radiation interaction is the formation of lattice defects, defect clusters, amorphous zones, dislocation loops or three-dimensional defects in inorganic material, polymerization, curing, grafting, chain secession, and crosslinking of organic material [6]. Surfaces damage can also take place in the form of adatoms, craters, and ripples or tracks [7]. Biological substances respond to radiation in a different manner, proteins, nucleic acids, cells, and multi-cell organisms -- very low doses (a few Gray) are sufficient to modify and inactivate organisms [8-10], while somewhat higher doses may impede structural studies although very high-energy gamma photons or electrons can also produce damage by atomic recoil processes [11] if the recoil energy exceeded the displacement threshold energy.

The present work aims to highlight the physics behind the interactions that occur at the microscopic scale -- above the common atomic-scale justification -- and the meaning of the common observations made on organic, inorganic, and biological substances; i.e. systematic effect on a specific property(ies) of the molecule/material/substance, some sort of dependence of the gamma radiation energy, and nonlinear dependence on the exposure dose. A mathematical model was proposed and discussed.

\noindent
\section{Mathematical model}

In the present work, we defined a term ``response function'' - notated ${\rm {\mathcal R}}_{\nu } $ - as the averaged signal obtained by the experiment(s) resulting of the measurement upon receiving a specific amount of radiation dose -- denoted ${\rm {\mathcal D}}$ - at dose rate $\dot{D}$. The subscript $\nu $ denotes the specific property measured in the investigation which reveals a measurement signal $I_{\nu } $. The measurement signals may be an infrared signature of a chemical species, crystal orientation or reorientation, survival curve of living microorganisms, electron spin resonance parameter or a radical formation, etc. The experiment(s) may be the relative, difference between signals of irradiated and un-irradiated substances, and may be absolute intensity obtained from an instrument for the irradiated substance. The fraction of sites affected by one unit dose at which the damage/change/modification of the property $\nu $ occurs (in suitable units such as Gy$^{-1} $) is denoted $\alpha _{\nu } $.  Considering a substance having a specific number $N_{o,v} $ of readily available sites per unit mass (or volume) at which the radiation makes its effect to obtain the specific signal $I_{\nu } $. The value of $\alpha _{\nu } $ is assumed to be constant, approximately, and proportional to the inverse of the specific number of readily available sites per unit mass (or volume), and it may be negative or positive. As the dose increases, these sites are exhausted according to the following equation:
\begin{equation} \label{GrindEQ__1}
\frac{dN_{\nu } }{N_{\nu } } \approx -\alpha _{\nu } \dot{D}dt+\beta _{\nu } \dot{D}dt-\delta _{\nu } \dot{D}dt
\end{equation}

Here, $\beta _{\nu } $ is a positive quantity representing the recovery factor so that the second term represents the recovery of radiation damage by any radiation repair processes [12-15]. The retardation enhancement factor, represented by the positive parameter $\delta _{\nu } ({\rm {\mathcal D}},t)$, the third term, which is a slowly varying function on the total dose received by the substance, the time elapsed since exposure, and dose rate at which it is exposed to radiation [16]; in the present work, we limit the mathematical problem on the dependence on the dose rate but we keep on mind that this factor has a long-term effect such as radiation aging and radical digestion. Eq. 1 can be solved as following over the exposure time ($\Delta t_{exp} $).
\begin{equation} \label{GrindEQ__2}
N_{\nu } =N_{o,\nu } \exp \left(-(\alpha _{\nu } -\beta _{\nu } +\delta _{\nu } )\dot{D}\Delta t_{\exp } \right)
\end{equation}
Or in terms of dose
\begin{equation} \label{GrindEQ__3}
N_{\nu } =N_{o,\nu } \exp \left(-(\alpha _{\nu } -\beta _{\nu } +\delta _{\nu } )\Delta {\rm {\mathcal D}}\right)
\end{equation}

The response function is proportional to the specific number of remained sites, upon integration with respect to increasing dose
\begin{equation} \label{GrindEQ__4}
{\rm {\mathcal R}}_{\nu } ({\rm {\mathcal D}})=\int _{0}^{{\rm {\mathcal D}}}N_{o,\nu } \exp \left(-(\alpha _{\nu } -\beta _{\nu } +\delta _{\nu } )x\right) \; dx
\end{equation}
Where $x$ is a dummy integration factor representing the fraction of dose received by the substance, i.e. response may change with dose as follows

\begin{equation} \label{GrindEQ__5}
{\rm {\mathcal R}}_{\nu } ({\rm {\mathcal D}})=\frac{N_{o,\nu } }{\alpha _{\nu } -\beta _{\nu } +\delta _{\nu } } \left(1-\exp \left(-(\alpha _{\nu } -\beta _{\nu } +\delta _{\nu } ){\rm {\mathcal D}}\right)\right),
\end{equation}
assigned for the property $\nu $. This signal function, on the other hand, depends on whether or not the measured property being part of the material before exposure. In other words, does this property signals the absolute existence of new synthesized species or the alteration made on that species after irradiation. In the first case, $\alpha _{\nu } $ is a positive quantity while in the second case $\alpha _{\nu } $ may be positive or negative depending on whether the species is generated or destroyed upon irradiation. In both cases, the signal is described by the difference between the final amount of sites and the original ones.
\begin{equation} \label{GrindEQ__6}
I_{\nu } =C_{1} \left({\rm {\mathcal R}}_{\nu } ({\rm {\mathcal D}})-{\rm {\mathcal I}}_{\nu } (0)\right)
\end{equation}
For the reason of simplification of the equation, the quantity $(\alpha _{\nu } -\beta _{\nu } +\delta _{\nu } )$ is replaced by $\alpha _{\nu }^{\bullet } $, where $\alpha _{\nu }^{\bullet } $ approaches --e.g. $10^{-12} $ Gy$^{-1} $ if the electrons are the affected sites. For DNA, for example, this fraction may reach $10^{-1} $ Gy$^{-1} $. That makes the quantity $\alpha _{\nu }^{\bullet } {\rm {\mathcal D}}$\textbf{ an exponential variable},which has a value between -1 and 1, as follow
\begin{equation} \label{GrindEQ__7}
I_{\nu } =C_{2} \left(1-\exp \left(-\alpha _{\nu }^{\bullet } {\rm {\mathcal D}}\right)\right)-C_{1} {\rm {\mathcal I}}_{\nu } (0)
\end{equation}
Where $C_{1} $and $C_{2} $ are the proportionality constants and ${\rm {\mathcal I}}_{\nu } (0)$is the response function of the pristine substance (un-irradiated material). The values of $C_{1} $and $C_{2} $ change sign if the property exists and deteriorated upon irradiation. Eq. 7  can be expanded by Taylor series as follows:
\begin{equation*} \label{GrindEQ__7a}
I_{\nu } =C_{2} \left(-\sum_{i=1}^{\infty} \frac{(-1)^i(\alpha _{\nu }^{\bullet } {\rm {\mathcal D}})^i}{i!}\right)-C_{1} {\rm {\mathcal I}}_{\nu } (0)
\end{equation*}

\begin{equation*} \label{GrindEQ__7b}
I_{\nu } =C_{2} \left(\alpha _{\nu }^{\bullet } {\rm {\mathcal D}} + \sum_{i=2}^{\infty} \frac{(-1)^{i+1}(\alpha _{\nu }^{\bullet } {\rm {\mathcal D}})^i}{i!}\right)-C_{1} {\rm {\mathcal I}}_{\nu } (0)
\end{equation*}

\noindent For small values of $|\alpha _{\nu }^{\bullet } {\rm {\mathcal D}}|$, the summation has a negligible value and the relation seems linear with dose. However, once the value become larger, higher terms become effective, Eq. 7 could be written as

\begin{widetext}
\begin{equation*} \label{GrindEQ__8a}
{{I}_{\nu }}={{C}_{2}}\left[ \sum\limits_{i=1}^{\infty }{{{(-1)}^{i+1}}\frac{{{(\alpha _{\nu }^{\bullet }\mathcal{D})}^{i}}}{i}}-\sum\limits_{i=1}^{\infty }{{{(-1)}^{i+1}}{{(\alpha _{\nu }^{\bullet }\mathcal{D})}^{i}}\left( \frac{1}{i}-\frac{1}{i!} \right)} \right]-{{C}_{1}}{{\mathcal{I}}_{\nu }}(0)
\end{equation*}
or
\begin{equation} \label{GrindEQ__8}
{{I}_{\nu }}={{C}_{2}}\left[ \ln\left(1 + \alpha _{\nu }^{\bullet }\mathcal{D}\right)-\sum\limits_{i=1}^{\infty }{{{(-1)}^{i+1}}{{(\alpha _{\nu }^{\bullet }\mathcal{D})}^{i}}\left( \frac{1}{i}-\frac{1}{i!} \right)} \right]-{{C}_{1}}{{\mathcal{I}}_{\nu }}(0)
\end{equation}
\end{widetext}
\noindent Here, the first summation is replaced by $\ln\left(1 + \alpha _{\nu }^{\bullet }\mathcal{D}\right)$, its the Taylor expansion. The second summation have first and second terms vanished. Higher order terms in the second summation, for $i\geq3$,  are negligible especially for values of $|\alpha _{\nu }^{\bullet }\mathcal{D}|$ less than 0.25. It can be neglected with 2\% bias from the real values. which is small compared to the measurement uncertainties in real cases; this would be extended practically to 0.5 with good precision as shall be shown in the discussion of Figs. 2 and 3. That means the special case of moderate dose is :
\begin{equation*} \label{GrindEQ__8a}
{{I}_{\nu }}\simeq{{C}_{2}}\ln\left(1 + \alpha _{\nu }^{\bullet }\mathcal{D}\right)-{{C}_{1}}{{\mathcal{I}}_{\nu }}(0)
\end{equation*}
\noindent provided that $|\alpha _{\nu }^{\bullet }\mathcal{D}| \leq 0.25$.

Finally at higher dose the full formulation of Eq. 8 should be used.
This expansion can be approximated by
\begin{equation} \label{GrindEQ__9}
{{I}_{\nu }}\cong\left\{ \begin{matrix} {{C}_{2}}\alpha _{\nu }^{\bullet }\mathcal{D}-{{C}_{1}}{{\mathcal{I}}_{\nu }}(0), & |\alpha _{\nu }^{\bullet }\mathcal{D}|<0.1  \\    \begin{array}{*{35}{l}} & {{C}_{2}}\ln (1+ \alpha _{\nu }^{\bullet }\mathcal{D})-{{C}_{1}}{{\mathcal{I}}_{\nu }}(0)  \\    {} & Eq.8  \\ \end{array} & \begin{array}{*{35}{l}},  & |\alpha _{\nu }^{\bullet }\mathcal{D}|\lesssim0.5  \\    {} & |\alpha _{\nu }^{\bullet }\mathcal{D}|\gtrsim0.5  \\ \end{array}  \\ \end{matrix} \right.
\end{equation}

For practical purposes of second line in Eq. 9, the deference on the response
\begin{equation} \label{GrindEQ__9}
({{I}_{\nu }}+C_1{{\mathcal{I}}_{\nu }}(0))/C_2\cong\left\{\begin{matrix} \alpha _{\nu }^{\bullet }\mathcal{D}, & |\alpha _{\nu }^{\bullet }\mathcal{D}|<0.1  \\    {{C}_{3}}\ln (\mathcal{D}), & |\alpha _{\nu }^{\bullet }\mathcal{D}|\lesssim0.5  \\ Eq. 8, & |\alpha _{\nu }^{\bullet }\mathcal{D}|\gtrsim0.5 \\ \end{matrix} \right.
\end{equation}
where $C_3$ is any experimentally determinable constant that is linearly proportional to the living fraction of sites affected by one unit dose at which the damage/change/modification of the property $\nu $ occurs, $\alpha _{\nu }^{\bullet }$.

\section{Discussion}

In previous report [6,17], it was emphasized that the primary interaction of gamma photon would not, by itself, induce an effect on the substance, except for few circumstances of the interaction with the atomic nuclei with an energy sufficient to induce atomic displacement or a photonuclear reaction. The reason was behind this was based on the fact that gamma interaction with matter is at the atomic scale of single-photon absorption with a short electromagnetic interaction time ($\tau = \hbar/E_\gamma \sim 10{}^{-21}$ s for ${}^{60}$Co gamma lines), while bond breakage requires interaction on longer timescales of order of femtoseconds [16]. The missing photo-electron or the Compton electron is, quickly recovered, by an electron from the molecular system or a lattice electron.

Because of the small probability of photo-nuclear interaction, the merely available courses for gamma interactions are those involving transfer of energy to an electron. Electron-nucleus interaction has also a small probability compared to Coulomb's interaction. Hence, even when using energetic photons we must quickly think in terms of the electrons. The emergence of secondary electrons is responsible for subsequent x-ray emission/reabsorption, ionization, and excitation of the surrounding atoms in the material, either in Markovian pattern and/or in cascade. The time elapsed between the states of PI, whatever are they, and the states of full acquisition of the energy by the bulk of the material are long enough to affect the thermodynamic properties of the substance within a limited volume (spores) around the PI.

The interaction of radiation with matter can be summarized in the time scale shown in Figure 1 adopted from Ref. [6] and lately from Ref. [18]. Primary interaction (PI) occurs in a very short time, the cascade excitation (CE) forms the electron cloud, a kind of plasma (P) that holds for a limited time. After a while, electron relaxation (ER) and lattice relaxation (LR) bring the material into thermodynamic (TD) equilibrium. In exceptional situations, collisional interactions displace atom(s) from their position leading to the formation of point-defect (PD) and dislocations loops (DL) inside the solid crystals.
\begin{figure}[h]
  \centering
  \includegraphics[width=\linewidth]{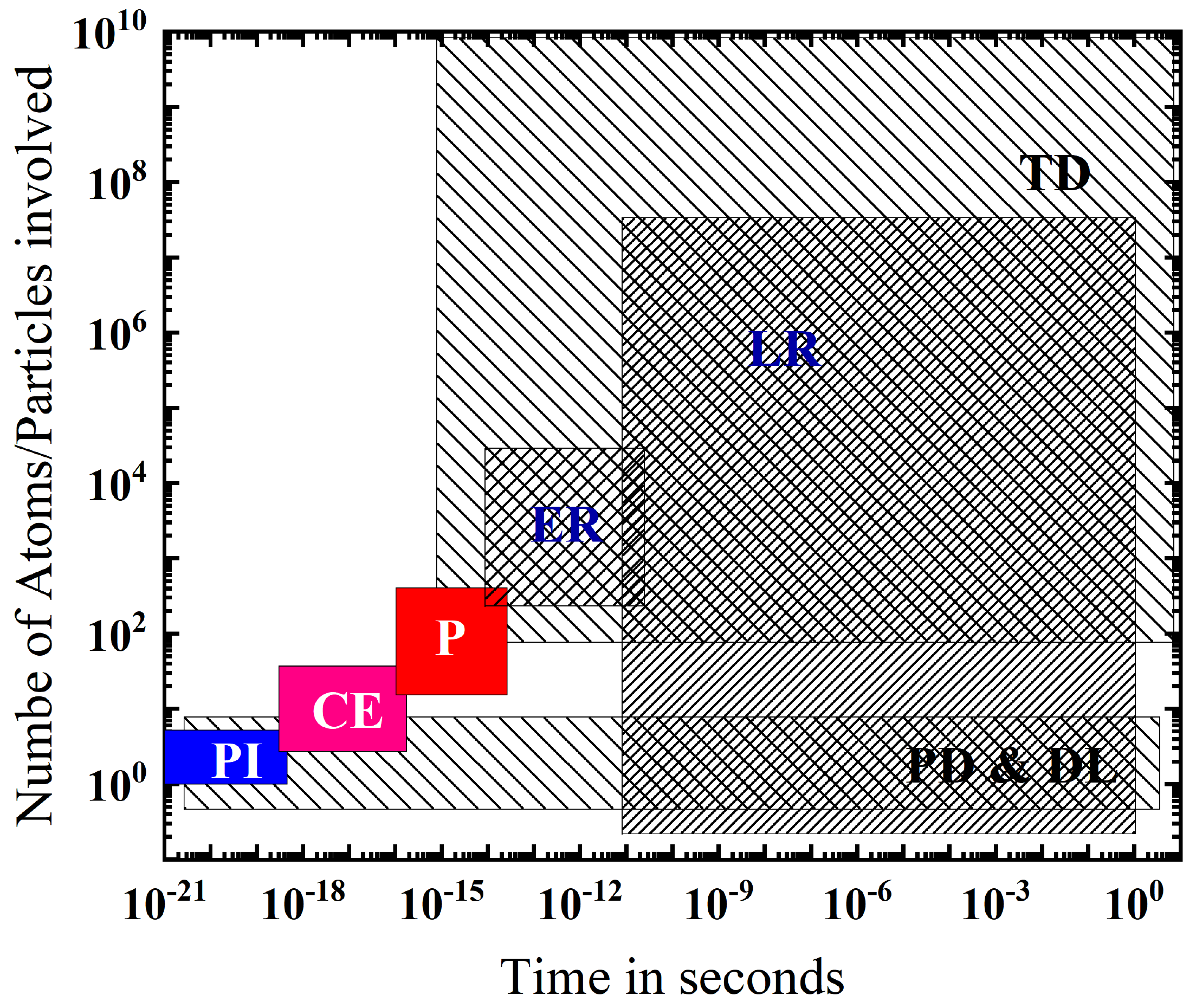}
  \caption{Time scale for the specific number of species (atoms/electrons) contributing to the effect of radiation on matter. PI represents the Primary Interaction, CE: the cascade excitation, P: the electrons cloud, ER: electron relaxation, LR lattice relaxation, TD thermodynamic equilibrium, PD point-defect and DL dislocations loops.  Adopted from [6,18].}\label{Fig1}
\end{figure}

Looking at the interaction of radiation with the matter as a single event will lead to inaccurate conclusions. For the moment, the specific number of available sites in Eq. 5 will be of the order of Avogadro's number while $\alpha _{\nu } $, which is proportional to the fraction of sites affected by one unit dose, became infinitesimally small compared to the specific number of sites. If PIs are random, and if the phenomenon investigated by the experimental technique was random, the response function of the physical system should be linearly variant with the received dose according to the first approximation in Eq. 9. This is illustrated as the shaded area for linear response in Fig. 2.
\begin{figure}
  \centering
  \includegraphics[width=\linewidth]{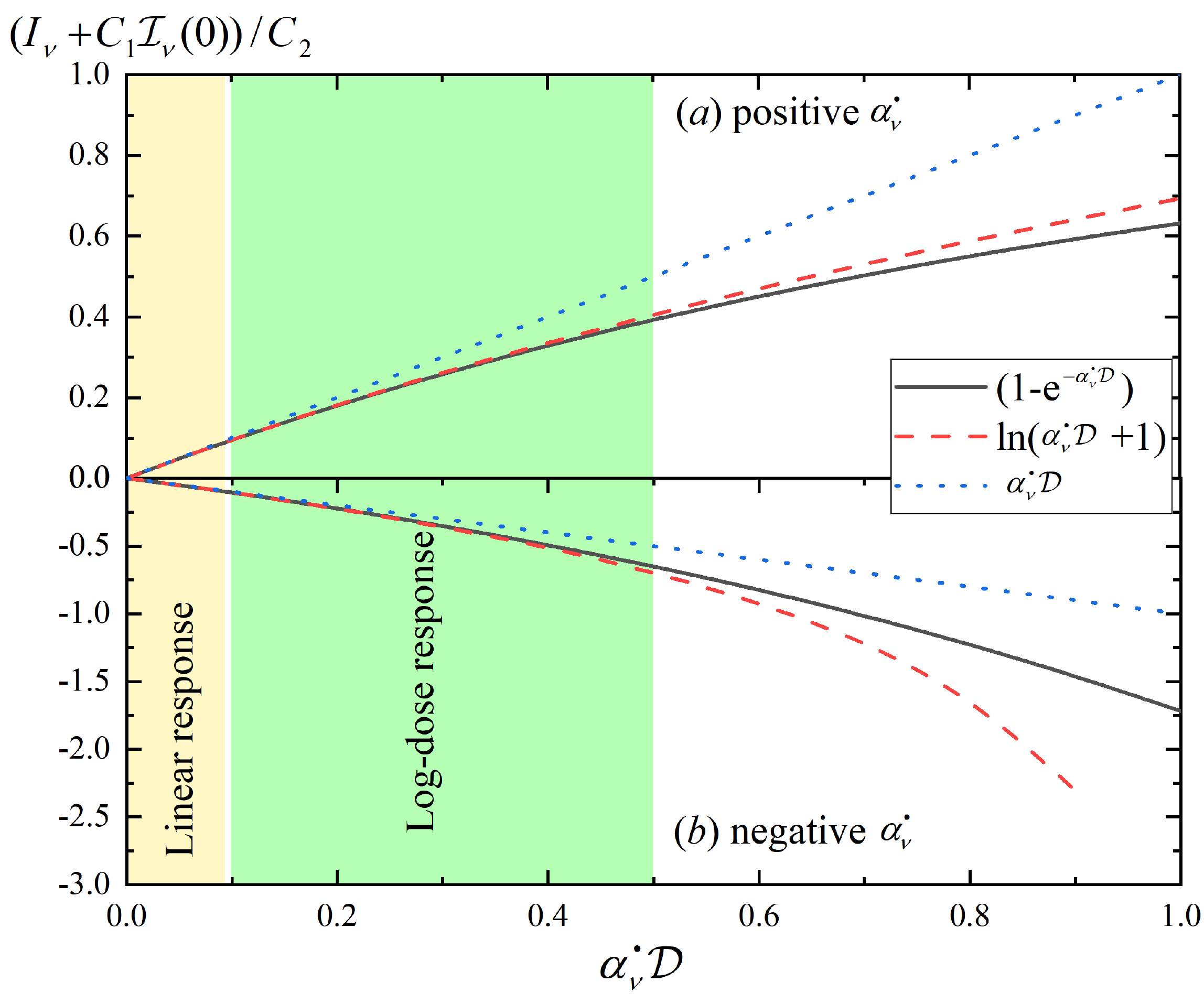}
  \caption{The variation of the signal function with dose (a) for positive value of $\alpha _{\nu }^{\bullet } {\rm {\mathcal D}}$, part (b) for its negative value of $\alpha _{\nu }^{\bullet } {\rm {\mathcal D}}$.}\label{fig2}
\end{figure}

\begin{figure}
  \centering
  \includegraphics[width=\linewidth]{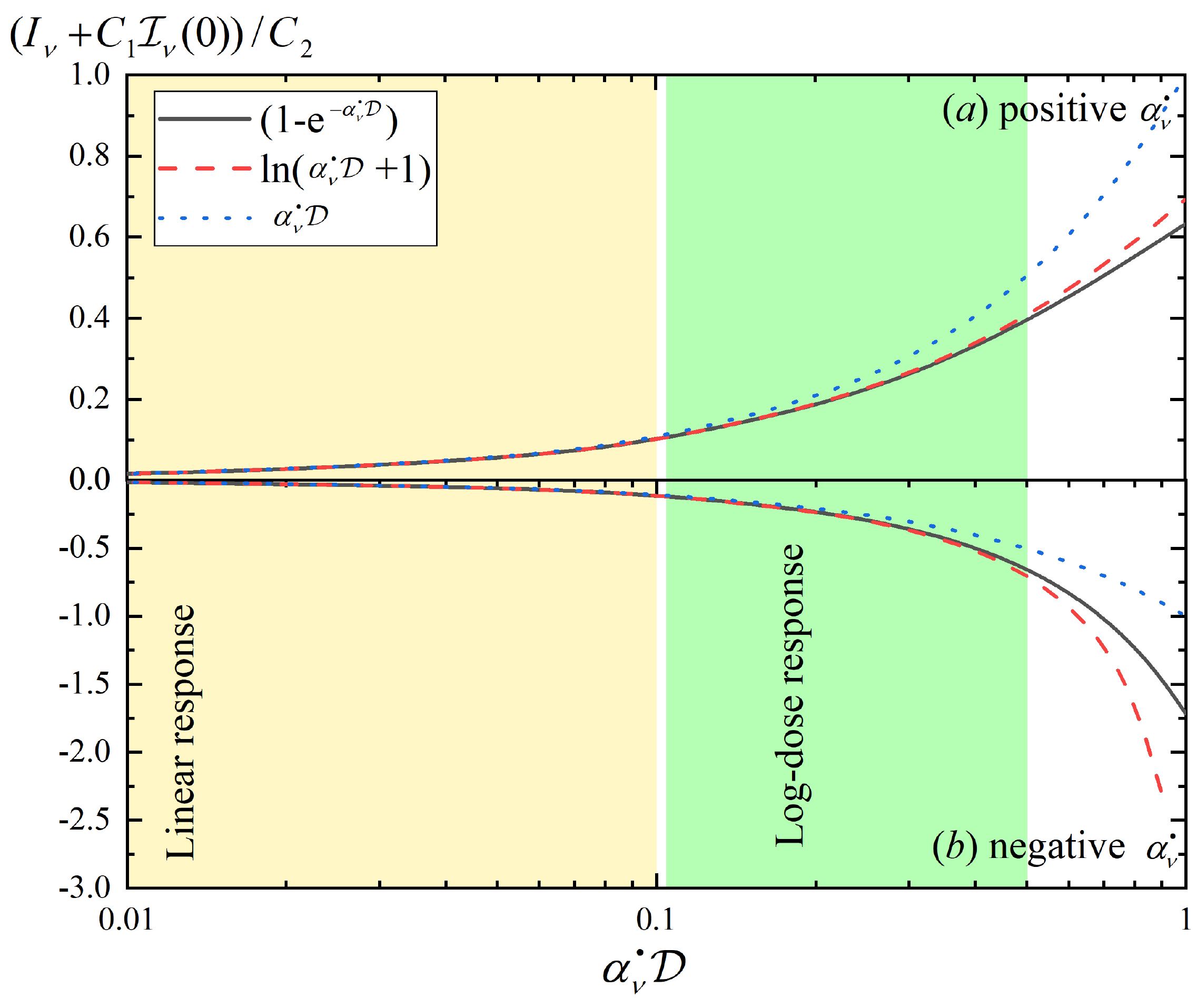}
  \caption{Illustration of log-Dose response, constructed as in Fig. 2.}\label{fig3}
\end{figure}

In a real situation, the PI produces a large number of secondary electrons which forms a spore of specific thermodynamic conditions that may change the property of the substance. What happened in practice is that specific portions of the molecule (for example) are affected by radiation, which is related to the thermodynamic of the interaction. That means, explicitly, if we have a property that signals the effect of radiation on a species of the material that has limited concentration per gram, the effect of the radiation shall be specific for the reaction around that site and the value of $\left|\alpha _{\nu }^{\bullet } \right|$ became larger. Literally speaking, the macroscopic dose is reflected on the material through the value of $\left|\alpha _{\nu }^{\bullet } \right|$.Within values of $\alpha _{\nu }^{\bullet } $ between - 0.5 and 0.5 the response function, and the signal upon radiation interaction varies as the second part of Eq. 9 in what is called log-dose response [6], see Fig. 3. Here, the nonlinearity induced by exhaustion of active sites shall affect our viewpoint concerning the radiation effect.

At higher doses, i.e.$\left|\alpha _{\nu }^{\bullet } {\rm {\mathcal D}}\right|>0.5$, the material toward the saturation of the change of the property. The log-dose response had been used by several researchers [15-17,19-25] to obtain a linear correlation between the measured property and the dose up to 5 MGy. In their researches, they used ($\log {\rm {\mathcal D}}$) instead of ${\rm {\mathcal D}}$\textbf{ }in their analysis of the effect.

\noindent
\section{Conclusions}

The effect of gamma- and electrons-radiation on organic, inorganic, and biological substances would not be considered as a result of primary interactions if the properties under investigation have a limited number of affected sites in the material

In biological substances, the specific number of effectible sites that affect the cell reproduction and normal metabolic functions are restricted to the region of the cell's nucleus within the DNA/RNA molecule while those sites that affect other cell organic components and organelles are relatively larger than those of DNA/RNA. Hence, if the property measured is related to the reproduction of cells and their normal metabolic functions, the effect of gamma-ray and electron radiation will be severe due to the large value of $\left|\alpha _{\nu }^{\bullet } \right|$. The value of $\left|\alpha _{\nu }^{\bullet } \right|$ may be reduced if the cell has DNA damage repair mechanisms due to the reverse sign of the recovery factor ($\beta _{\nu } $) as in Eq. 5; however, in case of RNA, the recovery factor ($\beta _{\nu } $) has smaller values and the process of damage is enhanced in the cell cytosol due to radical formation (which increases the enhancement factor, represented by $\delta _{\nu } $ in Eq. 5). RNA and other enzymes, proteins, and cell molecules as a part of the whole cell had a relatively larger number of affected sites with less efficient and specific repair mechanisms compared to DNA repair mechanisms and with larger enhancement factor; its corresponding radiation effect on the survival of the cell (survival analysis) will have a value of  $\left|\alpha _{\nu }^{\bullet } \right|$ comparable to those of DNA but for a different physical reason. Hence, radiation damages induced on biological substances are, in general, severe and reach value of $\left|\alpha _{\nu }^{\bullet } {\rm {\mathcal D}}\right|\approx 0.5$ at relativity small dose ( $<$$<$ 1 Sv).

None-living organic material especially synthesized materials are different from those biological substances in having a larger number of the specific number of effectible sites depending on the molecular weight of the organic substance and its molecular structure, physical form, and density. There is no repair mechanism in none-living organic molecules, but the re-polymerization and responding mechanism. However, it is difficult to restore the original structure of the polymer or the organic material. The signals that may be associated with radiation effects are infrared absorption, electron spin resonance, and structural morphology. These properties have small value of $\left|\alpha _{\nu }^{\bullet } \right|$ and may require higher dose to observe changes in such structures in response to radiation $>$ 1 KGy. For the change of hardiness properties in solid organic materials, only small number of deficiency in the molecular structure could make a large change in the hardiness properties, here a specific number of effectible sites is small and consequently, the signal function reaches a value of $\left|\alpha _{\nu }^{\bullet } {\rm {\mathcal D}}\right|\approx 0.5$ at relativity small dose ( \~{} 1 kGy). For property such as a change in the optical band-gap in response to radiation dose, the response function is summed over its component that affects the energy --gap such as supped over the doses required by each type of molecular bonding to break and summed over all other structural properties that affect the energy-gap. Routs of radiation synthesis of organic species from organic species have a large value of the specific number of effectible sites comparable to the Avogadro's number, which make a value of $\left|\alpha _{\nu }^{\bullet } \right|$ positive and very small. So, the relation, in this case, is nearly linear up to 5 MGy of dose.

The reason inorganic compounds are more resistive to radiation damage is their large value of the specific number of effectible sites comparable to the Avogadro's number, except for the cases in which radiation damages affects doping in the material and like Routs of radiation synthesis of organic species, its value of $\left|\alpha _{\nu }^{\bullet } \right|$ would be extremely small. A similar exception is made for the properties that include the formation of bubbles and the swelling of the material. However, if the property measured in the inorganic substance involves trapping of excitation, such as the case of thermoluminescent detectors (TLD), and/or electron-hole pair in semiconductor or similar signals that have a very limited room for registering the effect, the value of a specific number of effectible sites are restricted by the solid properties. In such cases, its value of $\left|\alpha _{\nu }^{\bullet } \right|$ is expected to be large and the linearity of the material's response became limited to a very small dose (${\rm \lesssim }$ 100 mSv). Increasing the dose above a certain limit will cause non-linearity (which is a common problem in radiation dosimetry).

\section*{Author contribution}

\noindent \textbf{Elsayed~K.~Elmaghraby}: Conceptualization, Methodology, Formal analysis, Writing - Original Draft.

\noindent \textbf{Mohamed Bahaaeldin-Afifi}: Software, Validation, Resources, Visualization, Data Curation.


\begin{thebibliography}{73}

\bibitem{1} Salaheldin, G. Assessment of radon radiological hazards in some ophiolite rocks, North Eastern Desert, Egypt. J Radioanal Nucl Chem  \textbf{328}, 447-454 (2021). https://doi.org/10.1007/s10967-021-07660-9

\bibitem{2} Eissa, H.S., Medhat, M.E., Said, S.A.  Elmaghraby, E. K.  Radiation dose estimation of sand samples collected from different Egyptian beaches, \textit{Radiation Protection Dosimetry}, \textbf{147(4)}, 533--540 (2011). https://doi.org/10.1093/rpd/ncq498

\bibitem{3} Ali, A., Elmaghraby, E.K. Detection and interference of fission-neutron reactions on third period elements, \textit{Nuclear Instruments and Methods in Physics Research Section B: Beam Interactions with Materials and Atoms}, \textbf{471}, 63-68 (2020). https://doi.org/10.1016/j.nimb.2020.03.028

\bibitem{4} Rudberg, E. The Energy Distribution of Electrons in the Photoelectric Effect, \textit{Phys. Rev.} \textbf{48}, 811 (1935). https://doi.org/10.1103/PhysRev.48.811

\bibitem{5} More, C.V., Alsayed, Z., Badawi, M.S., Thabet, A.Z., Pawar, P.P. Polymeric composite materials for radiation shielding: a review. \textit{Environ Chem Lett}, \textbf{19, }2057-2090\textbf{ }(2021). https://doi.org/10.1007/s10311-021-01189-9

\bibitem{6} Elmaghraby, E. K. \textit{Radiation interaction with matter: An approach at the nanometer scale. In: Radiation Synthesis of Materials and Compounds}, CRC Press, Taylor \& Francis Group (2013): 403-422.

\bibitem{7} Nordlund, K., Zinkle, S.J., Sand, A.E., Granberg, F., Averback, R.S., Stoller, R.E., Suzudo, T., Malerba, L., Banhart, F., Weber, W.J., Willaime, F., Dudarev, S.L., Simeone, D. Primary radiation damage: A review of current understanding and models, \textit{Journal of Nuclear Materials}, \textbf{512, }450-479 (2018). https://doi.org/10.1016/j.jnucmat.2018.10.027

\bibitem{8} Wu,Q., Allouch, A., Martins, I., Modjtahedi, N., Deutsch, E., Perfettini, J.-L. Macrophage biology plays a central role during ionizing radiation-elicited tumor response, \textit{Biomedical Journal}, \textbf{40(4)}, 200-211 (2017). https://doi.org/10.1016/j.bj.2017.06.003

\bibitem{9} Greenstock, C. Radiation and aging: free radical damage, biological response and possible antioxidant intervention. \textit{Med. Hypotheses} \textit{,}\textbf{ 41, }473-482 (1993).

\bibitem{10} Kempner, E. S. Effects of high-energy electrons and gamma rays directly on protein molecules. \textit{J. Pharm. Sci. }, \textbf{90}, 1637-1646 (2001).

\bibitem{11} Elmaghraby, E.K. Resonant neutron-induced atomic displacements, \textit{Nuclear Instruments and Methods in Physics Research Section B: Beam Interactions with Materials and Atoms}, \textbf{398}, 42-47 (2017). https://doi.org/10.1016/j.nimb.2017.03.054

\bibitem{12} Zibrov, M., Dürbeck, T., Egger, W., Mayer, M. High temperature recovery of radiation defects in tungsten and its effect on deuterium retention, \textit{Nuclear Materials and Energy}, \textbf{23}, 100747 (2020). https://doi.org/10.1016/j.nme.2020.100747

\bibitem{13} Chiba, A., Kawabata, N., Yamaguchi, M., Tokonami, S., Kashiwakura, I.  Regulation of Antioxidant Stress-Responsive Transcription Factor Nrf2 Target Gene in the Reduction of Radiation Damage by the Thrombocytopenia Drug Romiplostim, \textit{Biol.  Pharm.  Bull.} \textbf{43}, 1876--1883 (2020). https://doi.org/10.1248/bpb.b20-00442

\bibitem{14} Andrea M. Patterson, Tong Wu, Hui Lin Chua, Carol H. Sampson, Alexa Fisher, Pratibha Singh, Theresa A. Guise, Hailin Feng, Jessica Muldoon, Laura Wright, P. Artur Plett, Louis M. Pelus, Christie M. Orschell; Optimizing and Profiling Prostaglandin E2 as a Medical Countermeasure for the Hematopoietic Acute Radiation Syndrome. \textit{Radiat Res.} \textbf{195 (2)}, 115--127 (2021). https://doi.org/10.1667/RADE-20-00181.1

\bibitem{15} Elmaghraby, E.K., Seddik, U. Thermophysical properties and reaction kinetics of $\gamma $--irradiated poly allyl diglycol carbonates nuclear track detector, \textit{Radiation Effects and Defects in Solids}, \textbf{170(7-8)}, 621-629 (2015), https://doi.org/10.1080/10420150.2015.1077332

\bibitem{16} Abdelaal, S, Abdelhady, A.M., Tokhy, HH., Eid, A.M., Rammah, Y.S., Awad, E.M., Elmaghraby, E.K. Breeding behavior of radiation-induced effects in organic materials and their possible use as radiation dosimeters, \textit{Journal of Physics and Chemistry of Solids}, \textbf{150}, 109814 (2021). https://doi.org/10.1016/j.jpcs.2020.109814

\bibitem{17} Tokhy, HH, Elmaghraby, E. K., Abdelhady, A. M., Eid, A. M., Rammah, Y. S., Awad, E. M. and Abdelaal, S. The influence of gamma radiation on organic compounds having carbon ring and its application in dosimetry, \textit{Radiochimica Acta}, \textbf{109, }407-418 (2021). https://doi.org/10.1515/ract-2020-0024

\bibitem{18} Nordlund K., Short M.P. \textit{Modeling of Radiation Damage in Materials: Best Practices and Future Directions. In: Andreoni W., Yip S. (eds) Handbook of Materials Modeling}. Springer, Cham. (2020) pp. 2367-2379. https://doi.org/10.1007/978-3-319-44680-6\_146

\bibitem{19} Zaki, M.F., Elmaghraby, E.K. Photoluminescence of Gamma-Radiation Induced Defect on Poly AllylDiglycol Carbonates, \textit{J.} \textit{Luminescence}, \textbf{132(1)}, 119-121 (2012). http://dx.doi.org/10.1016/j.jlumin.2011.08.001

\bibitem{20} Salama, T.A. Elmaghraby, E.K. High Gamma-Ray Dose Measurement Using Nuclear Track Detector'', \textit{Radiation Protection and Dosimetry,} \textbf{140(3)}, 218--222 (2010). http://dx.doi.org/10.1093/rpd/ncq118

\bibitem{21} Elmaghraby, E.K., Salama, T.A.  Investigation of the fluorescence emitted from poly allyldiglycol carbonate modified by gamma-ray radiation excited by UV radiation, \textit{Radiation Effects and  Defects  in Solids}, \textbf{165 (4)}, 321--328 (2010). http://dx.doi.org/10.1080/10420150903452252

\bibitem{22} Zaki, M. F., Elmaghraby, E.K., Elbasaty, A.B. Structural alterations of polycarbonate/PBT by gamma irradiation for high technology applications, \textit{Journal of Adhesion Science and Technology}, \textbf{30(4)}, 443-457 (2016). http://dx.doi.org/10.1080/01694243.2015.1105123

\bibitem{23} Elmaghraby, E.K., Abdelaal,S., Abdelhady, A.M.,  Fares,S., Salama, S., Mansour, N.A. Experimental determination of the fission-neutron fluence-to-dose conversion factor'', \textit{Nuclear Instruments and Methods in Physics Research Section A}, \textbf{949}, 162889 (2020). https://doi.org/10.1016/j.nima.2019.162889

\bibitem{24} Elmaghraby, E.K., Abdelaal, S., Abdelhady,A.M. Fares, S., Safwat Salama, S. Mansour, N.A. Correspondence and difference between gamma-ray and neutron irradiation effects on organic materials in marine environment'', \textit{Egyptian Journal of Aquatic Biology \& Fisheries} \textbf{23(5)}, 1 -- 16 (2019). https://dx.doi.org/10.21608/ejabf.2019.63408

\bibitem{25} Elmaghraby, E.K. Abd El Aal, S., Awny, O., Mansour, N.A., Hassan, N.M., Salama, S.  Investigation of the reactor's high neutron flux effects on the physical and  chemical characteristics of polymeric material, \textit{Nuclear Inst. and Methods in Physics Research B} \textbf{461}, 210--218 (2019). https://doi.org/10.1016/j.nimb.2019.10.011
\end{thebibliography}
\end{document}